# Reconciling high resolution climate datasets using KrigR

Richard Davy[1*], Erik Kusch[2]

[1]Nansen Environmental and Remote Sensing Center, Jahnebakken 3, Bergen, Norway

[2]Center for Biodiversity Dynamics in a Changing World (BIOCHANGE), Section for

Ecoinformatics & Biodiversity, Department of Biology, Arhus University

*richard.davy@nersc.no

**ABSTRACT**

There is an increasing need for high spatial and temporal resolution climate data for the wide community of researchers interested in climate change and its consequences. Currently, there is a large mismatch between the spatial resolutions of global climate model and reanalysis datasets (at best around 0.25º and 0.1º respectively) and the resolutions needed by many end-users of these datasets, which are typically on the scale of 30 arcseconds (~900m). This need for improved spatial resolution in climate datasets has motivated several groups to statistically downscale various combinations of observational or reanalysis datasets. However, the variety of downscaling methods and inputs used makes it difficult to reconcile the resultant differences between these high-resolution datasets. Here we make use of the KrigR R-package to statistically downscale the world-leading ERA5(-Land) reanalysis data using kriging. We show that kriging can accurately recover spatial heterogeneity of climate data given strong relationships with co-variates; that by preserving the uncertainty associated with the statistical downscaling, one can investigate and account for confidence in high-resolution climate data; and that the statistical uncertainty provided by KrigR can explain much of the difference between widely used high resolution climate datasets (CHELSA, TerraClimate, and WorldClim2) depending on variable, timescale, and region. This demonstrates the advantages of using KrigR to generate customized high spatial and/or temporal resolution climate data.

## 1. INTRODUCTION

Ongoing climate change is having wide-reaching effects worldwide. This has fuelled a need for high spatial and temporal resolution climate data to be able to quantify and predict the effects of climate change (Hewitt et al. 2017; Trisos et al. 2020; Bjorkman et al. 2018). There has been a particular focus on creating high spatial resolution climate data products and several groups have now created products with resolutions as fine as 30 arcseconds (~900m) by statistically interpolating observations, reanalysis products, climate model outputs, or some combination thereof (Abatzoglou et al. 2018; Fick and Hijmans 2017; Karger et al. 2017; Beyer et al. 2020; Navarro-Racines et al. 2020). Currently, there exist several such datasets that offer unique configurations of variables, period covered, methodological and data background, and spatial and temporal resolution. However, due to the diversity of data sources and methods used, all of these high-resolution datasets contain numerically different climate data, particularly in locations with a low density of in-situ observations such as Alaska. This presents a serious challenge for users of these datasets who need to know which dataset they can trust for their particular purpose. Reconciling the differences between these different products is a difficult task which is exacerbated by the lack of uncertainty metrics associated with the high-resolution data. None of the existing high-resolution climate products account for the uncertainty in the underlying climate data, or in the downscaling technique, which can lead to over-confidence of end-users in the validity of these products. One tool that has emerged which may be used to address this challenge is KrigR (Kusch, Davy, *in prep.*).

KrigR is an R-Package which offers functionality for the retrieval and pre-processing of ERA5(-Land) data as well as statistical downscaling of spatial products to high spatial resolutions using kriging. A major advantage of this tool is that it preserves the uncertainty associated with the statistical downscaling and it gives access to the dynamical uncertainty information available from the ERA5 10-member ensemble of reanalyses. Here we make use of KrigR to downscale two of the most advanced global reanalyses, ERA5 and ERA5-Land. The ERA5 reanalysis has the unique advantage over other reanalyses in that it also includes a measure of the uncertainty associated with the climate data, which is derived from the use of ensemble data assimilation (Hersbach et al. 2020). This measure of data uncertainty is henceforth referred to as dynamical uncertainty. We demonstrate how the combination of statistical uncertainty (owing to the downscaling procedure) and dynamical uncertainty can explain much of the difference between widely used high resolution climate datasets including WorldClim2, TerraClimate, and CHELSA. Additionally, this exercise also highlights areas of potential concern of data accuracy of these same widely used high resolution climate datasets. This emphasizes the importance of preserving uncertainty due to statistical downscaling, so that it can then be incorporated into downstream applications of high-resolution climate data.

## 2. DATA & METHODS

All analyses were carried out in R(R Core Team; R Foundation for Statistical Computing; Vienna; Austria 2020) and MatLab (MATLAB 2018). Fully reproducible code for data acquisition, aggregation, and production of products for analyses within this study are available (see data availability statement).

### 2.1 DATA RETRIEVAL

Here we make use of the R-package KrigR to obtain and statistically downscale the ERA5 and ERA5-Land reanalysis products. These are the latest global climate reanalyses from the European Center for Medium-range Weather Forecasting (ECMWF) (Hersbach et al. 2020). Climate reanalyses products are the culmination of decades of research into data assimilation methodologies, dynamical models for the Earth system, and investment in Earth Observation (Buizza et al. 2018). Reanalyses use data assimilation to optimally combine a wide range of surface and satellite observations with a dynamical model in order to produce a self-consistent dataset which includes all essential climate variables (Sabater 2017; Hersbach et al. 2020). The ERA5 dataset includes a reanalysis and the underlying 10-member ensemble of forecasts used to derive the reanalysis. The spread in this 10-member ensemble gives us a measure of both the observational uncertainty (which is included in the data assimilation framework) and the stochastic uncertainty from the dynamical model. ERA5-Land (Sabater 2017) is a global land-surface reanalysis that was created by dynamically downscaling ERA5 to a resolution of 0.1º (11km).

For the analysis presented here we acquired the surface air temperature (SAT) and soil moisture from the first soil layer (Qsoil, 0-7cm depth) from ERA5-Land for the years 1981-2010 at hourly and monthly resolutions. The yearly resolution data were made by taking the annual-mean of the monthly data. We also acquired the ensemble members for SAT and Qsoil from ERA5 for the same period at both hourly and monthly resolutions. The ensemble spread was found by taking the standard deviation of the 10-member ensemble. This ensemble spread is referred to as the dynamical uncertainty. The average dynamical uncertainty shown in Figure 2C and 2F were calculated by taking the mean uncertainty at each temporal resolution across the period 1981-2010. Since it is extremely computationally expensive to downscale every timestep for the period 1981-2010, even at monthly resolution, we chose to take samples of the statistical uncertainty at each temporal resolution to establish the typical statistical uncertainty at a given temporal resolution. For the hourly data we averaged the statistical uncertainty from downscalings at 0000 and 1200 UTC for the 15$^{th}$ day of the months January, April, July, and October in 1981; for the monthly data we took the average statistical uncertainty for downscalings of monthly means for January, April, July and October in 1981; and for the yearly data we used the average statistical uncertainty from downscalings of the years 1981, 1991,

2001, and 2010. This was to control for diurnal, seasonal and interannual variations in the statistical uncertainty owing to changes in the strength of the statistical relationships between the target variable and co-variates. However, as can be seen from Figures 2C and 2F, there is very little variation in the statistical uncertainty across a wide range of timescales.

We acquired monthly SAT data from TerraClimate (Abatzoglou et al. 2018), WorldClim2 (Fick and Hijmans 2017), and CHELSA (Karger et al. 2017) for the period 1981-2000. This is the common period between these datasets. Each of these datasets are publicly available. See data availability statement for details. For TerraClimate (Abatzoglou et al. 2018) and WorldClim2 (Fick and Hijmans 2017) the climatologies of the diurnal-mean surface air temperature were made by taking the mean of the diurnal minimum and maximum temperatures, and then averaging over time. This averaging of the diurnal minimum and maximum temperatures to compute the diurnal-mean temperature is standard World Meteorological Organization protocol for station data (Thorne et al. 2016).

Elevation data was obtained through the KrigR package which acquires the data from the USGS GMTED 2010 open-access database (Danielson, J.J., Gesch 2011). We acquired datasets describing the slope steepness and slope aspect at 30 arcsecond resolution from the Harmonized World Soil Database v1.2 (Fischer et al. 2008). These were used on their native resolution, and also aggregated to the ERA5-Land resolution using the raster package (Hijmans and van Etten 2012). The soil thermal and hydrological parameters were obtained from the Land-Atmosphere Interaction Research group at Sun Yat-sen University (Dai et al. 2013). The soil parameters we use are the saturated water content, $\theta_s$, the saturated capillary potential, $\varphi_s$, the pore size distribution, $\lambda$, and the saturated hydraulic conductivity, $K_s$, from the Clapp and Hornberger functions; as well as the heat capacity of solid soils, $c_{soil}$, thermal conductivity of saturated soil, $\lambda_{sat}$, and the thermal conductivity for dry soil, $\lambda_{dry}$.

## 2.2 Statistical Interpolation

The KrigR package carries out statistical downscaling through kriging - a statistical interpolation technique. Kriging is a two-step process that requires training data that we wish to downscale, and co-variate data both at the resolution of the training data and at our target spatial resolution (Chilès and Delfiner 2012). In the first step, we fit variograms to our training data and establish covariance functions with our co-variate data at the training resolution. This gives us functions which describe the spatial autocorrelation of our training data, and its relationship with our chosen co-variate(s). During the second step we predict the value of our variable at new locations, in this case at a higher spatial resolution, using co-variate data at the target resolution. One major advantage to kriging is that it preserves the uncertainty obtained when fitting the variogram, which gives us an uncertainty associated with the downscaled data. In KrigR this uncertainty is given as a standard deviation of the uncertainty in the estimate (Hiemstra et al. 2009).

There are a few important limitations to all statistical downscaling methodologies, and for some variables there is simply no reasonable way to statistically downscale them. What cannot be accounted for when using any statistical downscaling approach is the effect of dynamical processes which occur only at the unresolved scales of the input data. For example, suppose one wishes to downscale temperature data from a resolution of 100 km to a resolution of 1 km. The dynamical model that was used to create the 100 km resolution data will account for how air temperature varies with altitude, because this can be partially resolved at these scales, but it will not include the effect of atmospheric circulation within a valley on air temperature. This can become important when the atmosphere is stably stratified and cold pools(Mutiibwa et al. 2015) or frost pockets can form in topographical depressions. For this reason, dynamical models of the atmosphere include descriptors of the un-resolved topography of the surface (e.g.

heterogeneity of slope angle) which are used to account for such unresolved processes. Whether or not statistical downscaling can account for these processes depends upon whether the underlying process is represented in the training data, and whether the relevant co-variates are used in the kriging. This essentially puts a limit on how large a change in resolution can be accomplished using statistical downscaling. As a general guide we do not advise downscaling to a resolution more than ten times finer than the training data – and so we have added a warning in KrigR to the user to caution against this.

For some variables, such as precipitation, the processes that determine their spatial pattern at finer resolutions than the training data are largely determined by atmospheric dynamics. Therefore, no combination of topographical co-variates is going to enable us to statistically downscale precipitation with high accuracy. We therefore do not recommend statistically downscaling precipitation data. However, there can be alternatives which also tell us about the water availability at high resolution, such as soil moisture, that we can successfully statistically downscale by using the soil properties and topographical properties as co-variates as we have done here.

### *2.3 Dynamical and Statistical Uncertainty*

KrigR provides statistical uncertainty alongside downscaled products. This statistical uncertainty is the uncertainty resulting from statistical downscaling and is given as a standard deviation of the uncertainty in the estimate, $\sigma_{Krig}$, at each timestep. The magnitude of this uncertainty will depend upon (1) the robustness of the relationships between the target variable and the co-variates, (2) the spatial variability of the training data, and (3) the change in resolution between training and target resolutions (Chilès and Delfiner 2012).

In addition to this statistical uncertainty, the ERA5 reanalysis provides uncertainty due to the dynamics of the climate system and the limited observations used to constrain the reanalysis, henceforth referred to as dynamical uncertainty (Hersbach et al. 2020). In ERA5, this uncertainty information is provided as the 10 individual members of the ensemble used to generate the reanalysis, and as a measure of the ensemble spread (standard deviation of the 10 members), which can both be acquired using KrigR. Here, we refer to this measure as is $\sigma_{Dyn}$. The magnitude of the dynamical uncertainty depends upon the coverage and accuracy of the observations and the sensitivity of the model physics, which both depend upon location and climate variable. For example, there is a large and accurate array of observations of surface air temperature (Osborn et al. 2021; Menne et al. 2018), but temperature is also very sensitive to local conditions especially in complex terrain (Mutiibwa et al. 2015). In contrast, soil moisture has relatively poor observations (Robock et al. 2000; Dorigo et al. 2011), but is less variable in time than surface air temperature.

To calculate the dynamical uncertainty at different temporal resolutions, as shown in Figure 2, we first average the hourly data from each of the 10 members of the ERA5 ensemble at the given temporal resolution, and then take the standard deviation of the ensemble. We then found the average dynamical uncertainty over the full period 1981-2000 by taking the mean of the variance. For example, for the annual data we took:

$$\overline{\sigma_{Dyn_{annual}}} = \sqrt{\frac{1}{20} \sum_{year=1981}^{2000} \sigma_{Dyn_{annual}(year)}^2}$$

Similarly, to find the average statistical uncertainty from multiple timesteps we also take the square root of the mean of the variance at each timestep. To calculate the total uncertainty, i.e. the combined dynamical and statistical uncertainty, we take the square root of the mean of the variances in the statistical and dynamical uncertainty:

$$\sigma_{Total} = \sqrt{\frac{\sigma_{Krig}^2 + \sigma_{Dyn}^2}{2}}$$

Finally, to calculate the spread in the difference between the datasets at each timestep we first calculated the differences between the datasets at each timestep e.g. for CHELSA this is given by: $T' = T_{Krig} - T_{CHELSA}$. We then fitted a normal distribution to these differences which gave us both the mean difference between the datasets $\mu_{Err}$ and the spread in the differences, $\sigma_{Err}$, for each gridcell.

## 3. RESULTS

### *3.1 CHOICE OF CO-VARIATES IS IMPORTANT FOR DOWNSCALING AND UNCERTAINTY*

The most important factor that may affect the results of statistical downscaling is the choice of co-variate(s) (Chilès and Delfiner 2012). The most important variables affecting the interpolation of surface climate fields are the descriptors of the surface topography: elevation, slope angle, and slope steepness (Daly et al. 2002). For example, elevation is strongly related to the surface temperature field as air temperatures tend to decrease strongly with altitude (typically by 6.5 K km$^{-1}$). Furthermore, the direction which sloping terrain is facing and its steepness can also strongly affect surface air temperature by altering the amount of downwelling solar radiation per unit area that is absorbed at the surface. Slope direction and steepness can also strongly affect runoff, and hence soil moisture content.

Figure 1 shows the effect of using different co-variates when kriging surface air temperature and soil moisture over the UK at a randomly chosen monthly time-step. We chose the UK as an example due to large variations in topography as well as soil properties. Our downscaled product shows that elevation is extremely important in determining the surface air temperature and that the spatial pattern thereof closely matches that of elevation (Figure 1A). Adding the soil thermal properties (Cressie 1988) (see Data and Methods) of the terrain as a co-variate (Figure 1B) changes the estimate of local temperature by up to 0.4K across the UK. The effect of accounting for soil thermal properties is especially pronounced in mountainous regions such as the Scottish Highlands where there is a large degree of heterogeneity in soil properties. Including the slope steepness to the elevation-driven kriging (Figure 1C) also changes the estimate, but to a lesser degree with local changes of up to 0.1K. Doing so reveals clear patterns in how the estimate changes, especially in the complex terrain of the Scottish Highlands: there is a general warming at high altitudes and cooling in valleys. This pattern is due to the co-variance between elevation and slope steepness resulting in some component of the vertical temperature gradient being assigned to steepness instead of elevation. By including slope aspect with elevation (Figure 1D), the estimate of surface air temperature changes by around 0.4K for select areas. However, in this case, there are less clear spatial patterns to how the estimate changes as slope orientations can change completely from one grid-cell to the next. For soil moisture, also shown in Figure 1 (E-H), the picture is very different. Accounting for the soil properties (Figure 1E) is important, but the addition of elevation (Figure 1F) has little effect on the downscaling output. Introducing slope steepness (Figure 1G) also had little effect. However, including slope aspect (Figure 1H) as a co-variate can change the estimate by up to 0.1 kg kg$^{-1}$, a change of more than 15% from the estimate using soil properties alone. This tells us that, on monthly-averaged timescales, soil moisture is strongly determined by the soil thermal and hydrological properties. It is also more determined by the direction the terrain is facing, and thus the amount of solar radiation it receives than by how steep the slope is or how high the elevation. Note that the strength of these relationships is likely to vary with temporal resolution, time of year, and geographical region.

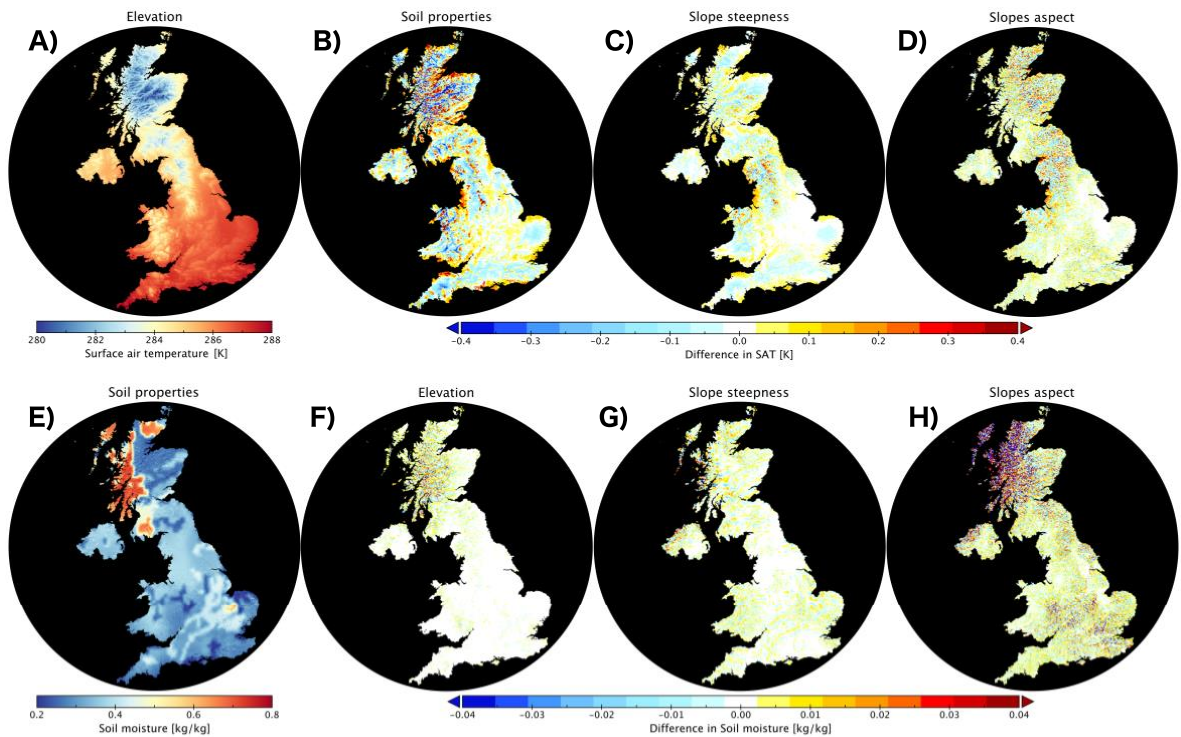

*Figure 1. **The effect of different Co-Variates on downscaling.** **A)** The surface air temperature for a single monthly mean downscaled from ERA5-Land at a resolution of 0.1º (11km) to a resolution of 30 arcseconds (~900m) using elevation as the only co-variate, and the difference in the estimate of the downscaled temperature when we add **B)** Soil thermal and hydrological properties, **C)** Slope steepness, and **D)** Slope aspect as co-variates. **E)** The soil moisture for a single monthly-mean downscaled from ERA5-Land to a resolution of 30 arcseconds (~900m) using soil thermal and hydrological properties as the only co-variates, and the difference in the estimate of the downscaled soil moisture when we add **F)** Elevation, **G)** Slope steepness, and **H)** Slope aspect as co-variates.*

### 3.2 DYNAMICAL UNCERTAINTY IS COMPARABLE TO STATISTICAL UNCERTAINTY

Uncertainties change with timescale. The magnitude of dynamical uncertainty can change from hour to hour, but at hourly resolutions the dynamical uncertainty of surface air temperature can be of the same magnitude or larger than the statistical uncertainty. Figure 2 contrasts the dynamical and statistical uncertainty for surface air temperature and soil moisture across the UK for a snapshot at a randomly chosen hour (A, B, D, E) and at different temporal resolutions for the period of Jan/1981-Dec/2010 (C, F). The similarity in magnitude of dynamical and statistical uncertainty at hourly resolutions can be seen in both the snapshot (A, B) and the average uncertainty (C, F). However, the dynamical uncertainty of surface air temperature decreases rapidly with longer timescales, while the statistical uncertainty remains similar across a range of timescales from hourly to 30-year climatologies (C). This consistency in statistical uncertainty implies that the strength of the relationship between surface air temperature and elevation is very similar across this range of timescales. For soil moisture, the dynamical uncertainty is always small relative to the statistical uncertainty, even at hourly resolutions (F). Just like with air temperature, dynamical uncertainty of soil moisture also has large spatial and temporal variability whereas statistical uncertainty is near-constant in space and time.

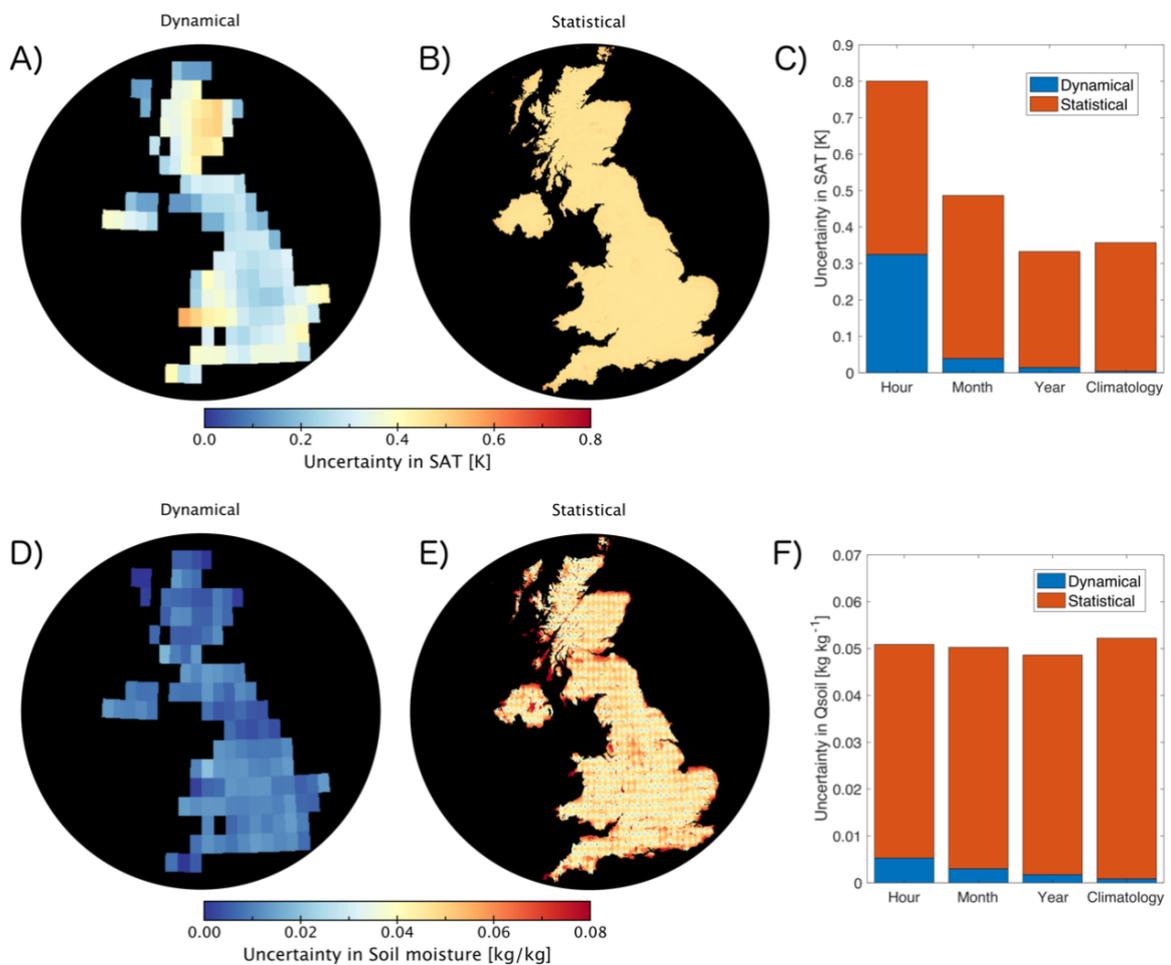

*Figure 2. **Dynamical vs. Statistical Uncertainty.** A) The dynamical uncertainty of the SAT for a single hour from ERA5, **B**) the statistical uncertainty in the SAT obtained from KrigR when ERA5-Land data for the same*

*hour as in A) is downscaled to a resolution of 30 arcseconds (~900m) using elevation as co-variate, **C**) the average dynamical and statistical uncertainty in SAT across the UK domain for hourly, monthly, yearly, and climatological-mean temporal resolutions for the period 1981-2010. **D**), **E**), and **F**) show the same for soil moisture except that soil thermal and hydrological properties are used as the co-variates for downscaling as in Figure 1E.*

While dynamical uncertainty can be comparable in magnitude to statistical uncertainty neither can or should be used as a proxy for the other.

### *3.3 Kriging recovers spatial variability*

Given the absence of independent, high resolution climate data with which to measure downscaling skill of KrigR, we assess how well kriging can recover spatial variability from coarse-grained ERA5-Land climate variables. First, we aggregate the ERA5-Land variable to a coarser resolution (0.4°; 40km), then we use KrigR to statistically downscale the data back to the original ERA5-Land resolution (0.1°; 11km). We then compare the results of the kriging against two other commonly used interpolation techniques: nearest neighbour and bilinear interpolation. The results are summarised in Figure 3. Figure 3A shows a box plot of the mean absolute difference between the downscaled SAT and the original (i.e. upscaled) data. Nearest neighbour interpolation is the simplest approach and so gives us a useful point of reference. We see nearest-neighbour interpolation produces the largest differences, with bilinear interpolation giving us slightly better results. The best interpolation is the kriging using just elevation as a co-variate. When we start to add other co-variates, the estimates get further from the original data. This illustrates the strength of the relationship between SAT and elevation and gives us confidence in downscaling SAT even from climate model projections which typically have resolutions of around 0.5°. Furthermore, KrigR provides the uncertainty of the estimated SAT in the form of a standard deviation of the uncertainty in the estimate ($\sigma_{Krig}$). This was combined with the dynamical uncertainty to create a total uncertainty, $\sigma_{Total}$. We compared this total uncertainty to the error in our estimated SAT by fitting a normal distribution to the differences between the kriged SAT and the original ERA5-Land data for all months in the period 1981-2010 to determine the spread in the downscaling error ($\sigma_{Err}$). Figure 3B shows the normalized difference between these two uncertainties. In all but a few coastal locations the uncertainty given by KrigR is larger than the actual spread in the errors; meaning that in almost all cases the difference between our downscaled data and the original ERA5-Land data lies within the uncertainty from KrigR.

The picture is a little different for Qsoil. In Figure 3C we can see that a simple nearest-neighbour interpolation actually gives us the best results. Furthermore, kriging with any of the covariates we considered actually produces worse results than a simple bilinear interpolation. However, the big advantage with kriging is that we also get an uncertainty associated with our estimate, and from Figure 3D we can see that the difference between the downscaling with kriging and the original ERA5-Land data is always within the uncertainty given by KrigR. So even in the case we do not have useful co-variates for our chosen downscaling, we can be confident that the uncertainty from the kriging captures the real uncertainty in our downscaling i.e., that the actual value at our target resolution, i.e. from the original ERA5-Land data, lies within the statistical uncertainty given by KrigR.

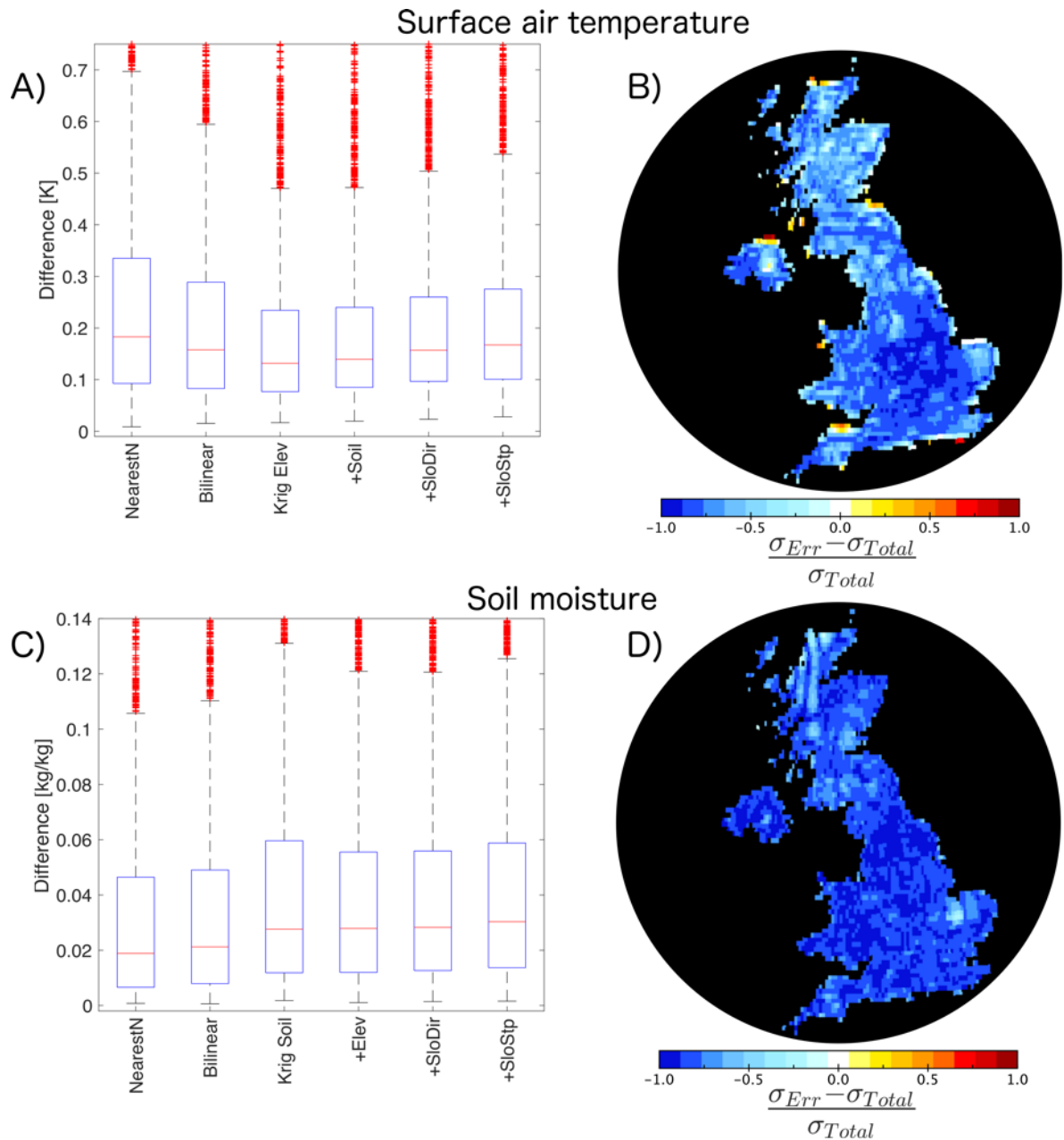

*Figure 3. **Downscaling Uncertainty and Confidence in Kriged Products.** A) Boxplots of the mean absolute difference between the downscaled surface air temperature and the original ERA5-Land data using 6 different interpolation techniques: nearest neighbour, bilinear, and four examples of kriging with different combinations of co-variates. B) The normalized difference between the spread in the monthly surface air temperature for the period 1981-2010 downscaled using kriging with elevation as a co-variate ($\sigma_{Err}$) and the mean of the standard deviation in the estimate from kriging ($\sigma_{Krig}$). C) and D) show the same for soil moisture.*

### 3.4 Trade-off with local kriging

321  One of the choices in KrigR is whether and how to localise the kriging. This defines the number
322  of neighbouring gridcells (nmax) used to derive the relationships between the field to be
323  downscaled and the co-variates. The larger the choice of nmax, the larger the number of cells
324  that will be used in the kriging process, and the closer the relationship between the target
325  variable and the co-variates becomes to the domain-average relationship. If nmax is set too
326  small then the relationships with the co-variates may be spurious, in which case so too will be
327  the downscaled product. This is represented in the fact that the smaller the nmax, the larger the
328  uncertainty in the downscaled product. This effectively enables users to fine-tune their choice
329  of nmax for individual needs and requirements within the cost-benefit trade-off between
330  computational resources and data uncertainty. Figure 4 shows a comparison where the same
331  monthly-mean surface air temperature data from ERA5-Land was downscaled to a 30
332  arcsecond resolution (~900m) using a range of choices for nmax from 15 to 480. These values
333  of nmax have been converted to an effective radius, given the resolution of ERA5-Land (0.1°,
334  11km). Here we have taken the downscaling using an nmax of 480 as the reference point. As
335  we go to decrease nmax, the mean absolute difference between the downscaled products
336  increases non-linearly, as does the uncertainty. However, the computational time needed to
337  downscale the data increases exponentially with increasing nmax. Herein lies the trade-off in
338  local kriging: Higher nmax values lead to convergence on the estimate for the downscaled
339  product and reduce the uncertainty, but dramatically increase the computational cost. This
340  raises the question of what nmax (and hence effective radius) to choose in local kriging. If we
341  consider meteorological variables, then the optimum effective radius for local kriging should
342  be similar to the scale of weather systems – about 100km – however, the optimum solution will
343  depend upon what level of accuracy is needed for the downscaled product. Furthermore, the
344  uncertainty in the original climate product due to the limited observations can be more
345  important than this uncertainty from the statistical downscaling, depending upon the variable and timescale of interest (see *3.2 Dynamical uncertainty is comparable to statistical uncertainty*).

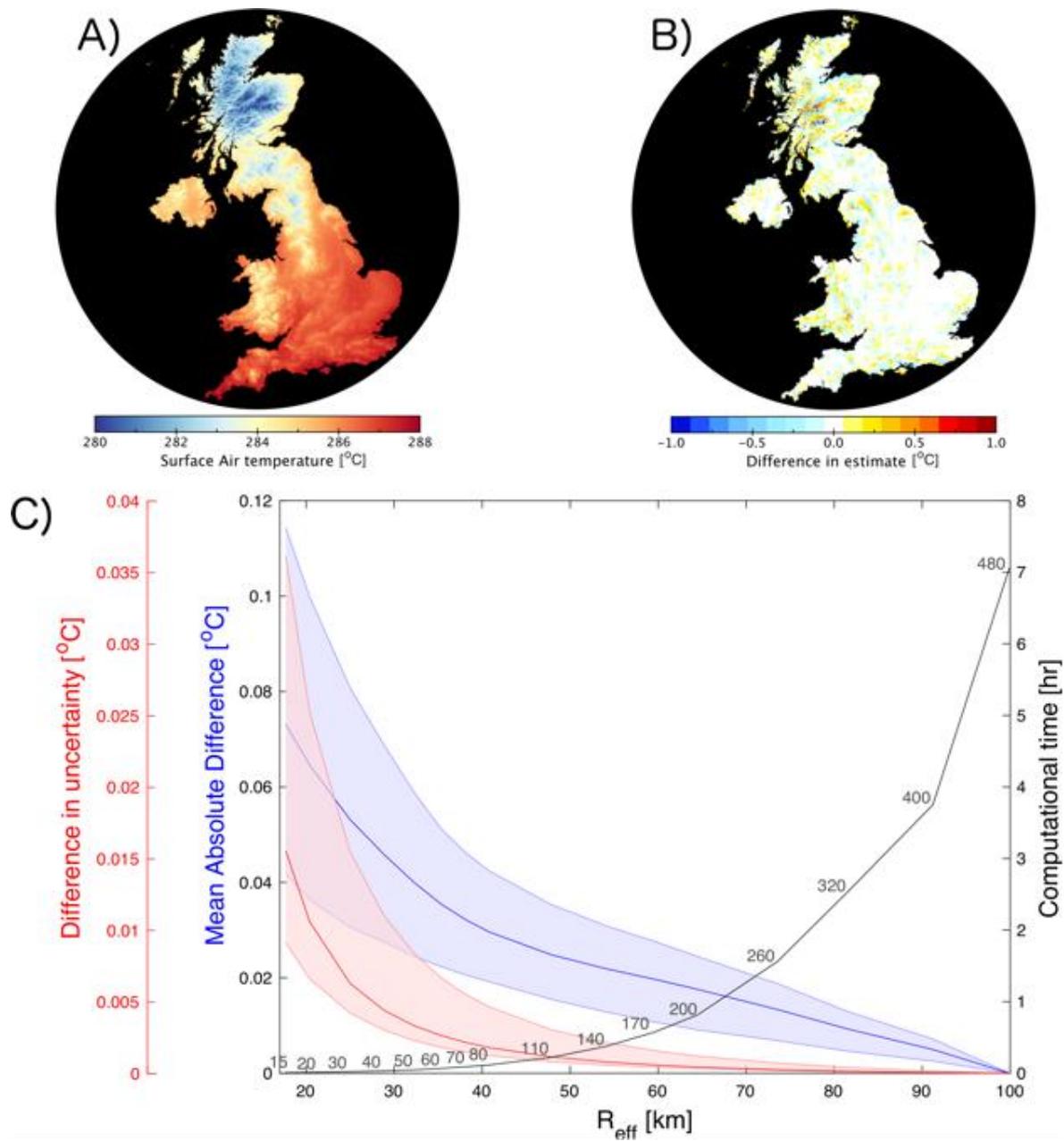

*Figure 4. **Localised Kriging and the nmax-trade-off.** A) The surface air temperature at monthly resolution downscaled from ERA5-Land at a resolution of 0.1° (11km) to a resolution of 30 arcseconds (~900m) using an nmax of 480. B) The difference in the surface air temperature downscaled using an nmax of 480 and 15. C) The mean absolute difference in the SAT (blue), the difference in the uncertainty in the downscaling (red), and the computational time needed for the downscaling (black) as a function of the effective radius of the local kriging, derived from nmax in the range 15 to 480 (grey numbers). The mean absolute difference and the difference in*

*the uncertainty are calculated using the nmax of 480 as the reference data. The thick lines show the mean value for the domain and the shaded areas mark the range from the 25$^{th}$ to the 75$^{th}$ percentiles.*

### *3.5 KrigR uncertainty explains much of the differences between high-resolution climate products*

There are several high-spatial-resolution climate data products already widely used in various scientific communities including WorldClim2 (Fick and Hijmans 2017), TerraClimate (Abatzoglou et al. 2018) and CHELSA (Karger et al. 2017) (see Table 1). Each of these datasets was created using different approaches and data sources. Therefore, there can be substantial differences between them. Climate products derived via KrigR have the advantage that they also contain the uncertainty (statistical and dynamical when queried) associated with the high-resolution climate data. We first tested KrigR-downscaling capabilities and accuracy using coarse-grained ERA5-Land data and demonstrated that the difference between our downscaled products and the original data always lie within the uncertainty around the downscaling predictions given by KrigR (see *3.3 Kriging recovers spatial variability*). We therefore evaluated the differences between the aforementioned high-spatial-resolution climate products to determine if these are also within the uncertainty given by KrigR. Since we might expect that this depends upon the density of observations used in creating these products, we assess this not just for the UK, but also for Alaska which has a very sparse ground-based observation network (Bieniek et al. 2014). Figure 5 shows the comparison between the combined dynamical and statistical ("total") uncertainty obtained via KrigR and the spread in the differences in monthly SAT between downscaled ERA5-Land data using KrigR and each of WorldClim2, TerraClimate, and CHELSA, respectively. Figures 5A and 5B show that the differences between downscaled ERA5-Land and TerraClimate and CHELSA over the UK are within our total uncertainty from KrigR everywhere except in mountainous regions such as the Highlands of Scotland. However, WorldClim2 lies well outside of the total uncertainty obtained from KrigR across the entire UK. So, by accounting for all uncertainty using KrigR, we can explain the differences between each of these products, except for WorldClim2. The fact that WorldClim2 is significantly different from each of these other high-resolution gridded products over the UK is most-likely due to the limitations of the data source – it was created using raw station data (Fick and Hijmans 2017). These station data are also included in the ERA5(-Land) reanalysis products (Hersbach et al. 2020), but in the data assimilation procedure there are numerous modifications made to the raw station data to account for changes to measurement procedures, devices used, and other aspects which affect the representivity of an individual station.

390  *Table 1. **Contemporary Climate Data Sets**. A comparison of contemporary high spatial resolution climate data sets which are widely used in analyses of climate impacts.*

| Name | Time-Period | Resolution | | Number of Parameters available | Area within KrigR uncertainty [%] | | Source |
|---|---|---|---|---|---|---|---|
| | | Spatial | Temporal | | UK | Alaska | |
| WorldClim 2.1 Climatologies | 1960-2018 | ~900m | 59 years | 26[1,2] | 82[3] | 59[3] | https://www.worldclim.org/data/worldclim21.html |
| WorldClim Historical monthly weather data | 1960-2018 | 21km | 1 month | 3 | 00 | 00 | https://www.worldclim.org/data/monthlywth.html |
| TerraClimate | 1958-2019 | 16km | 1 month | 14 | 58 | 02 | http://www.climatologylab.org/terraclimate.html |
| CHELSA | 1979-2013 | ~900m | 1 month | 46[1] | 72 | 16 | https://chelsa-climate.org/timeseries/ |

391

392  [1] 19 of these are bioclimatic variables which are derivatives of air temperature and water availability.

393  [2] 1 of these is elevation data.

394  [3] These are defined as the percentage of grid points where the difference in the climatologies is less than 1 standard deviation of the uncertainty,

395  normalised to the expected value of 68%.

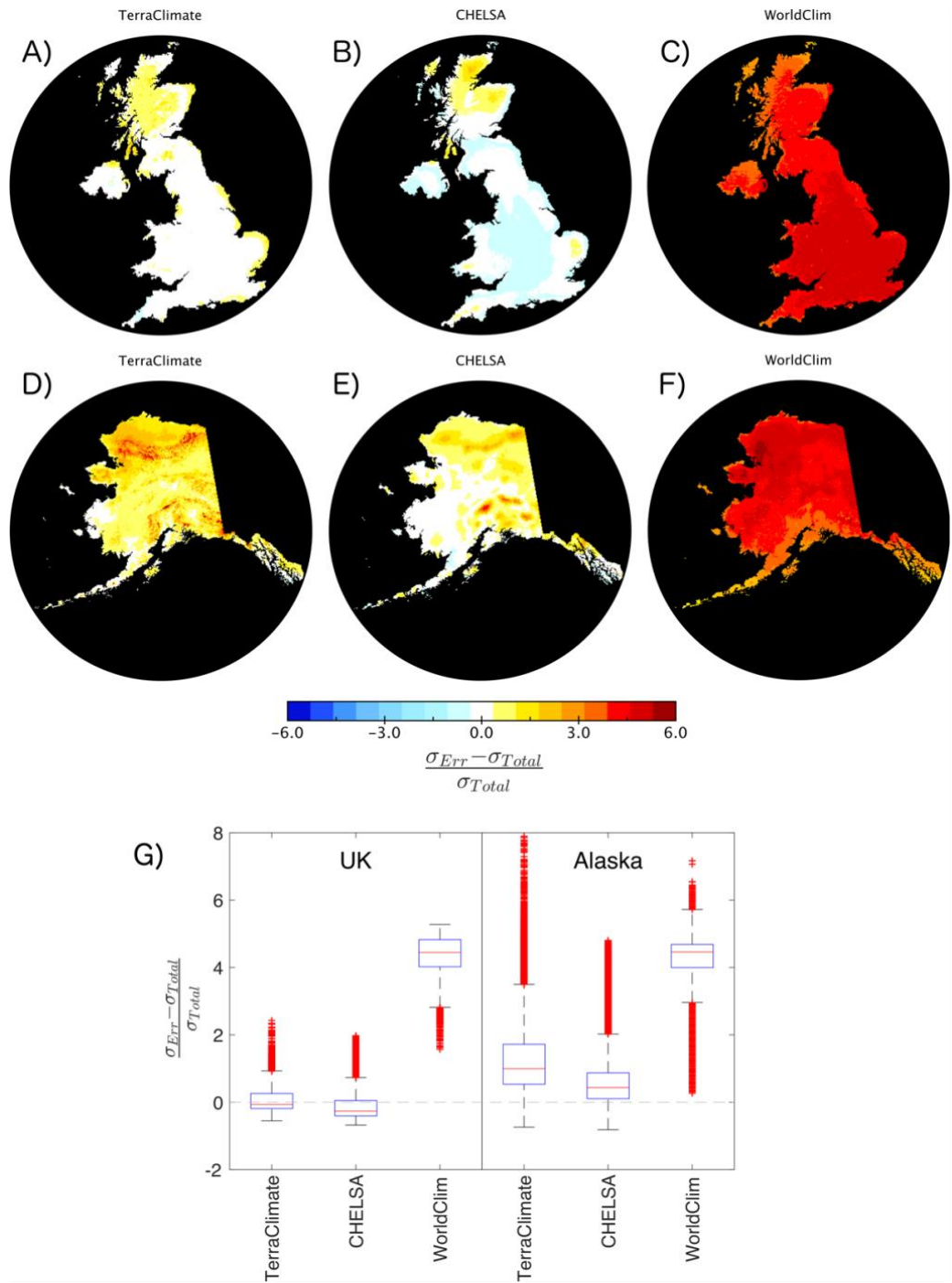

*Figure 5. **Kriged Products vs. Competitor Climate Products for all months Jan/1981-Dec/2010.** The difference between the standard deviation in the error between downscaled ERA-5Land SAT and **A)** TerraClimate, **B)** CHELSA, and **C)** WorldClim2 ($\sigma_{Err}$) and the standard deviation in the combined statistical and dynamical uncertainty from KrigR ($\sigma_{Total}$) normalised to this combined uncertainty from KrigR for the UK. These were calculated using monthly data for the period Jan/1981 to Dec/2010. **D), E),** and **F)** show the same for Alaska. Blue or white indicates that the dataset lies within the KrigR uncertainty, whereas yellow and red indicate that*

*the differences between the datasets are larger than expected from the KrigR uncertainty. **G)** shows the same data as a boxplot.*

There are much larger differences between our downscaled ERA5-Land and TerraClimate and CHELSA over Alaska than across the well-sampled UK, mostly due to large differences in the mountainous regions. However, WorldClim2 again stands out as being very different from the other three climate products and lies well outside of the uncertainty given by KrigR. Indeed, if we compare the boxplots in Figure 5G, it is apparent that there are larger differences between WorldClim2 and the other products in the well-observed climate of the UK than there is between ERA5-Land, CHELSA and TerraClimate in the poorly observed climate of Alaska. In the absence of independent weather-station data it is impossible to make judgements as to which data products to rely upon in areas where differences in KrigR-downscaling and other contemporary data sets are not explained by KrigR-uncertainty (i.e. yellow or red areas in Figure 5). Nevertheless, it is important to point out the advancements in data accuracy and temporal resolution of the ERA5(-Land) data family over those products used to build the aforementioned high-resolution climate data sets. Furthermore, the highly advanced kriging methodology used in KrigR regularly outperforms those interpolation methods that were used in the creation of other high-resolution climate data sets (see *3.3 Kriging recovers spatial variability*). Consequently, we show that climate products obtained via KrigR outperform competing, currently available climate products particularly in regions of topographic heterogeneity in terms of (1) temporal resolution, (2) spatial resolution (only ever matched by CHELSA and WorldClim2 climatologies, see table 1), and (3) accuracy.

## 4. CONCLUSIONS

The various high resolution climate data products currently available are an important source of climate information for diverse research and modelling communities including those in ecology and hydrology. However, the discrepancies in the climate data between these products force users to arbitrarily choose a single product for a given study since the datasets are inconsistent. Without a large set of independent, high-resolution observations against which to validate, it is impossible to determine the optimum product to use for a given purpose. However, it is crucial to use an accurate climate dataset when creating a high-resolution data product, and the ERA5(-Land) reanalysis has been demonstrated to outperform other global observation and reanalysis datasets for several key climate variables including the surface energy balance. Therefore, because of the demonstrated data accuracy and unrivalled temporal resolution, we recommend the adoption of ERA5(-Land) products as sources of environmental information, and as the basis for creating high-resolution climate datasets.

KrigR provides an R-integrated workflow for retrieval of ERA5(-Land) data as well as statistical interpolation capabilities to overcome mismatches in spatial resolution between data products. We have demonstrated that the methodology contained in KrigR enables R-users to create high-resolution climate data sets fit for individual study requirements that is of high spatial and temporal resolution, accurate, and accounts for uncertainty.

We have shown that by preserving the uncertainty associated with statistical downscaling, as well as that derived from observational datasets, the products created by KrigR can explain a large part of the difference between existing high-resolution climate data products, depending upon timescale, similarity of input data, and spatial resolution (Table 1). This also emphasizes the need to include measures of uncertainty in downstream applications and highlights areas of particular concern with respect to data accuracy.

We recommend that the current use of climate data products, particularly high spatial resolution products, for research and applications may need to be re-evaluated for the development of best-practice workflows. We expect efforts like KrigR to be a key steppingstone in reconciling high resolution climate data products and streamlining the choice/creation of appropriate data products for individual study needs.

## AUTHORS' CONTRIBUTIONS

E.K. and R.D. created all R scripts necessary for the analyses, and R.D. created all MatLab scripts necessary for the analyses. R.D. led the analyses presented here and created the figures. All authors contributed critically to the drafts and gave final approval for publication.

## CONFLICT OF INTEREST

The authors declare no conflict of interest associated with this work.

## DATA AVAILABILITY

All data used here are freely and publicly available. ERA5(-Land) data come from the European Center for Medium range Weather Forecasting (cds.climate.copernicus.eu). The digital elevation model data is available at the United States Geological Survey website (usgs.gov/centers/eros/science/usgs-eros-archive-digital-elevation-global-multi-resolution-terrain-elevation). The soil thermal and hydrological properties were obtained from globalchange.bnu.edu.cn/research/soil4.jsp and the slope aspect and steepness data are from the Harmonized World Soil Database v1.2 (fao.org/soils-portal/data-hub/soil-maps-and-databases/harmonized-world-soil-database-v12/en/). TerraClimate data can be acquired from climatologylab.org/terraclimate.html; WorldClim2 data from worldclim.org/; and CHELSA data from chelsa-climate.org.
Fully reproducible R code to obtain all data used within this study can be found here: https://github.com/ErikKusch/KrigRMS.

## REFERENCES

Abatzoglou, J. T., S. Z. Dobrowski, S. A. Parks, and K. C. Hegewisch, 2018: TerraClimate, a high-resolution global dataset of monthly climate and climatic water balance from 1958-2015. *Sci. Data*, **5**, 1–12, https://doi.org/10.1038/sdata.2017.191.

Beyer, R. M., M. Krapp, and A. Manica, 2020: High-resolution terrestrial climate, bioclimate and vegetation for the last 120,000 years. *Sci. Data*, **7**, 1–9, https://doi.org/10.1038/s41597-020-0552-1.

Bieniek, P. A., J. E. Walsh, R. L. Thoman, and U. S. Bhatt, 2014: Using climate divisions to analyze variations and trends in Alaska temperature and precipitation. *J. Clim.*, **27**, 2800–2818, https://doi.org/10.1175/JCLI-D-13-00342.1.

Bjorkman, A. D., and Coauthors, 2018: Plant functional trait change across a warming tundra biome. *Nature*, **562**, 57–62, https://doi.org/10.1038/s41586-018-0563-7.

Buizza, R., and Coauthors, 2018: The EU-FP7 ERA-CLIM2 project contribution to advancing science and production of earth system climate reanalyses. *Bull. Am. Meteorol. Soc.*, **99**, 1003–1014, https://doi.org/10.1175/BAMS-D-17-0199.1.

Chilès, J. P., and P. Delfiner, 2012: *Geostatistics: Modeling Spatial Uncertainty: Second Edition*. John Wiley & Sons, 1–699 pp.

Cressie, N., 1988: Spatial prediction and ordinary kriging. *Math. Geol.*, **20**, 405–421, https://doi.org/10.1007/BF00892986.

Dai, Y., W. Shangguan, Q. Duan, B. Liu, S. Fu, and G. Niu, 2013: Development of a china dataset of soil hydraulic parameters using pedotransfer functions for land surface modeling. *J. Hydrometeorol.*, **14**, 869–887, https://doi.org/10.1175/JHM-D-12-0149.1.

Daly, C., W. P. Gibson, G. H. Taylor, G. L. Johnson, and P. Pasteris, 2002: A knowledge-based approach to the statistical mapping of climate. *Clim. Res.*, **22**, 99–113,

496       https://doi.org/10.3354/cr022099.

497   Danielson, J.J., Gesch, D. B., 2011: Global Multi-resolution Terrain Elevation Data 2010
498       (GMTED2010). *U.S. Geol. Surv. Open-File Rep. 2011-1073*, **2010**, 26.

499   Dorigo, W. A., and Coauthors, 2011: The International Soil Moisture Network: A data hosting
500       facility for global in situ soil moisture measurements. *Hydrol. Earth Syst. Sci.*, **15**, 1675–
501       1698, https://doi.org/10.5194/hess-15-1675-2011.

502   Fick, S. E., and R. J. Hijmans, 2017: WorldClim 2: new 1-km spatial resolution climate
503       surfaces for global land areas. *Int. J. Climatol.*, **37**, 4302–4315,
504       https://doi.org/10.1002/joc.5086.

505   Fischer, G., F. O. Nachtergaele, S. Prieler, E. Teixeira, G. Tóth, H. van Velthuizen, L. Verelst,
506       and D. Wiberg, 2008: Global Agro-ecological Zones (GAEZ v3.0). *IIASA FAO*, 196.

507   Hersbach, H., and Coauthors, 2020: The ERA5 global reanalysis. *Q. J. R. Meteorol. Soc.*, **146**,
508       1999–2049, https://doi.org/10.1002/qj.3803.

509   Hewitt, C. D., R. C. Stone, and A. B. Tait, 2017: Improving the use of climate information in
510       decision-making. *Nat. Clim. Chang.*, **7**, 614–616, https://doi.org/10.1038/nclimate3378.

511   Hiemstra, P. H., E. J. Pebesma, C. J. W. Twenhöfel, and G. B. M. Heuvelink, 2009: Real-time
512       automatic interpolation of ambient gamma dose rates from the Dutch radioactivity
513       monitoring network. *Comput. Geosci.*, **35**, 1711–1721,
514       https://doi.org/10.1016/j.cageo.2008.10.011.

515   Hijmans, R. J., and J. van Etten, 2012: raster: Geographic analysis and modeling with raster
516       data.

517   Karger, D. N., and Coauthors, 2017: Climatologies at high resolution for the earth's land surface areas. *Sci. Data*, **4**, 1–20, https://doi.org/10.1038/sdata.2017.122.

MATLAB, 2018: *9.7.0.1190202 (R2019b)*. The MathWorks Inc.,.

Menne, M. J., C. N. Williams, B. E. Gleason, J. Jared Rennie, and J. H. Lawrimore, 2018: The Global Historical Climatology Network Monthly Temperature Dataset, Version 4. *J. Clim.*, **31**, 9835–9854, https://doi.org/10.1175/JCLI-D-18-0094.1.

Mutiibwa, D., S. Strachan, and T. Albright, 2015: Land Surface Temperature and Surface Air Temperature in Complex Terrain. *IEEE J. Sel. Top. Appl. Earth Obs. Remote Sens.*, **8**, 4762–4774, https://doi.org/10.1109/JSTARS.2015.2468594.

Navarro-Racines, C., J. Tarapues, P. Thornton, A. Jarvis, and J. Ramirez-Villegas, 2020: High-resolution and bias-corrected CMIP5 projections for climate change impact assessments. *Sci. Data*, **7**, 1–14, https://doi.org/10.1038/s41597-019-0343-8.

Osborn, T. J., P. D. Jones, D. H. Lister, C. P. Morice, I. R. Simpson, J. P. Winn, E. Hogan, and I. C. Harris, 2021: Land Surface Air Temperature Variations Across the Globe Updated to 2019: The CRUTEM5 Data Set. *J. Geophys. Res. Atmos.*, **126**, https://doi.org/10.1029/2019JD032352.

R Core Team; R Foundation for Statistical Computing; Vienna; Austria, 2020: R: A language and environment for statistical computing.

Robock, A., K. Y. Vinnikov, G. Srinivasan, J. K. Entin, S. E. Hollinger, N. A. Speranskaya, S. Liu, and A. Namkhai, 2000: The Global Soil Moisture Data Bank. *Bull. Am. Meteorol. Soc.*, **81**, 1281–1299, https://doi.org/10.1175/1520-0477(2000)081<1281:TGSMDB>2.3.CO;2.

Sabater, J. M., 2017: ERA5-Land: A new state-of- the-art Global Land Surface Reanalysis

540 Dataset.

541 Thorne, P. W., and Coauthors, 2016: Journal of Geophysical Research : Atmospheres. 5115–
542     5137, https://doi.org/10.1002/2015JD024583.Received.

543 Trisos, C. H., C. Merow, and A. L. Pigot, 2020: The projected timing of abrupt ecological
544     disruption from climate change. *Nature*, **580**, 496–501, https://doi.org/10.1038/s41586-
545     020-2189-9.

546